\documentclass[conference,a4paper]{IEEEtran}

\usepackage[ruled, vlined]{algorithm2e}
\usepackage{colortbl,epsfig}
\usepackage{comment}
\usepackage{cite}
\usepackage{enumerate}
\usepackage{subfigure}
\usepackage[normalem]{ulem}
\usepackage{url}
\usepackage{times}
\usepackage{multirow,multicol}
\usepackage{tabularx}
\usepackage{fixltx2e} 
\usepackage{graphicx}
\usepackage{psfrag}
\usepackage{amsfonts,amssymb,amsopn,amsmath,pifont,mathrsfs,accents}
\usepackage[mathscr]{eucal}
\usepackage{flushend}

\newtheorem{theorem}{Theorem}
\newtheorem{lemma}{Lemma}
\newtheorem{definition}{Definition}

\newcommand {\ab}	{\boldsymbol{a}}

\newcommand {\eb}	{\boldsymbol{e}}
\newcommand {\hb}  	{\boldsymbol{h}}
\newcommand {\pb}  	{\boldsymbol{p}}
\newcommand {\rb}  	{\boldsymbol{r}}
\newcommand {\ssb} 	{\boldsymbol{s}}
\newcommand {\vb} 	{\boldsymbol{v}}
\newcommand {\xb}	{\boldsymbol{x}}
\newcommand {\yb} 	{\boldsymbol{y}}
\newcommand {\Hm} 	{\mathbb{H}}
\newcommand {\Pm} 	{\mathbb{P}}
\newcommand {\const}	[1]{\mathscr{#1}} 
\newcommand {\Top}	{\textsf{T}}
\newcommand  {\C}     {\mbox{${\textstyle \,{\cal C}\!\!\!\!\!\!\!\sim}$}}

\newcounter{step}

\begin{document}

\title{Maximum-likelihood Soft-decision Decoding for Binary Linear Block Codes Based on Their Supercodes}

\author{
   \IEEEauthorblockN{
     Yunghsiang S. Han\IEEEauthorrefmark{1},
     Hung-Ta~Pai\IEEEauthorrefmark{2}, 
     Po-Ning~Chen\IEEEauthorrefmark{3} and 
Ting-Yi~Wu\IEEEauthorrefmark{3}
      }
   \IEEEauthorblockA{
     \IEEEauthorrefmark{1}Dep. of Electrical Eng.
    National Taiwan University of Science and Technology,
    Taipei, Taiwan\\
    Email: yshan@mail.ntust.edu.tw}
   \IEEEauthorblockA{
     \IEEEauthorrefmark{2}Dep. of Communication Eng.
   National Taipei University,
    Taipei, Taiwan}
   \IEEEauthorblockA{
     \IEEEauthorrefmark{3}Dep. of Electrical and Computer Eng.,
     National Chiao Tung University,
    Hsinchu, Taiwan}
  }

\maketitle

\begin{abstract}
Based on the notion of supercodes,
we propose a two-phase maximum-likelihood soft-decision decoding (tpMLSD) algorithm
for binary linear block codes in this work. 
The first phase applies the Viterbi algorithm backwardly to a trellis
derived from the parity-check matrix of the supercode of the linear block code.
Using the information retained from the first phase,
the second phase employs the priority-first search algorithm to 
the trellis corresponding to the linear block code itself, 
which guarantees finding the ML decision.
Simulations on  Reed-Muller codes 
show that the proposed two-phase scheme 
is an order of magnitude more efficient 
in average decoding complexity than the recursive maximum-likelihood decoding (RMLD) 
\cite{FUJ98} when the signal-to-noise ratio per information bit is $4.5$ dB.
\end{abstract}

%

\section{Introduction}
\pagestyle{plain}
Linear block codes have been deployed for error control in communication systems for many years, while algebraic structures of such codes are generally used for their decoding~\cite{SHU04}. Since the inputs of algebraic decoders are commonly required to be quantized into two levels, they are classified as hard-decision decoding technique.
In comparison with soft-decision decoding technique, 
a loss of information is induced due to quantization and hence
the decoding performance is restricted.

By contrast, the soft-decision decoding is developed to  eliminate the performance loss due to quantization. The input of soft-decision decoding is thus unquantized (or
practically quantized into more than two levels). In the literature, many maximum-likelihood (ML) soft-decision decoding algorithms for linear block codes have been proposed~\cite{WOL78,SNY91,LOU93,HAN93,KAN94,KAN97,GAZ97,FUJ98,AGU98,HAN981},
and the priority-first search algorithm (PFSA) is one of them~\cite{HAN93}.
It has been shown in \cite{HAN93} that the PFSA can provide the optimal ML decoding performance within practically acceptable decoding complexity.

In this paper, a novel two-phase maximum-likelihood soft-decision decoding (tpMLSD) scheme based on supercodes of linear block codes is proposed. Specifically, in the first phase, the Viterbi algorithm (VA)
is applied to a trellis, derived from the parity-check matrix of the supercode of the linear block code to be decoded, in a backward fashion (i.e., operated from the last trellis level to the first trellis level). Upon the completion of the first phase, each state
will retain a path metric that is used later in the second phase.
Because the trellis derived from the parity-check matrix of the supercode of the linear block code has fewer states than that derived from the parity-check matrix of the linear block code itself, the computational complexity is considerably reduced. 

In the second phase, the priority-first search algorithm is applied to the trellis corresponding to the parity-check matrix of the linear block  code. With a properly designed evaluation function, the optimal ML decision is guaranteed to be located. Notably, the path metric information obtained from the first phase 
is incorporated into the evaluation function for priority-first search, by which
the decoding procedure can be significantly sped up.
Simulations on Reed-Muller codes are then performed to confirm the efficiency   
of the proposed two-phase ML soft-decision decoding scheme.

It should be pointed out that the idea of decoding linear block codes based on their super codes is not new in the literature. It has been used in the hard-decision decoding in~\cite{BAR99},
where super codes are designed based upon covering sets and split syndromes.
In addition, a suboptimal hard-decision list decoding of linear block codes based on trellises of supercodes was presented in~\cite{FRE03}.
Further generalization of \cite{FRE03} to ML soft-decision list decoding 
 and to soft-output decoding can 
be found in \cite{FRE04} and \cite{FRE12}, respectively.

The rest of this paper is organized as follows. 
Notions of supercodes and ML soft-decision decoding of linear block codes
are introduced in Section~\ref{sec:ml}. 
The proposed two-phase ML soft-decision decoding algorithm for binary linear block codes
based on their supercodes is presented in Section~\ref{sec:proposed}. 
The optimality of the proposed algorithm is proved 
in Section~\ref{sec:optimality}. 
Section~\ref{sec:performance} evaluates the complexity of the proposed algorithm for practical linear block codes,
and Section~\ref{sec:conclusion} concludes the paper.

\section{Notions of Supercodes and ML Soft-decision Decoding of Linear Block Codes}
\label{sec:ml}
Let $\C$ be an $(n,k)$  binary linear block code with parity-check matrix $\Hm$ of size $(n-k)\times n$. Denote by $\overline\C$ an $(n,\bar k)$ supercode of $\C$ with parity-check matrix $\overline\Hm$ of size $(n-\bar k)\times n$, satisfying that
\begin{equation}
\label{eq:H}
\Hm=\begin{bmatrix}
\ \overline\Hm\ \ \\
\Pm
\end{bmatrix}
\end{equation}
for some matrix $\Pm$ of size $(\bar k-k)\times n$, where $\bar k>k$.

A trellis corresponding to linear block code $\C$ can then be constructed below.
 Denote by $\hb_j$, $0\le
j\le n-1$, the $(j+1)$th column of $\Hm$.  Let $\vb=
(v_0,v_1,\ldots,v_{n-1})$ denote a codeword of \C. 
By defining recursively a sequence of states 
$\{\ssb_\ell\}_{\ell=-1}^{n-1}$ as:
$$\ssb_\ell=\begin{cases}
{\bf 0},&\ell=-1\\
\ssb_{\ell-1}+v_\ell\hb_\ell,&\ell=0,1,\ldots,n-1,
\end{cases}
$$
a path corresponding to codeword $\vb$ on a trellis $\const{T}$ of $(n+1)$ levels
can be identified, where ${\bf 0}$ is the all-zero vector of proper size.
Obviously,   
$$\ssb_\ell=\sum_{j=0}^\ell v_j\hb_j\ \ \text{ for }\ell=0,1,\ldots,n-1$$
and
$$\ssb_{n-1}={\bf 0}\ \ \text{ for all codewords of \C}.$$
The trellis $\const{T}$ derived from $\Hm$ is then formed by 
picking up all paths corresponding to codewords of \C.

By convention,  state
$\ssb_\ell$ identifies a node on trellis $\const{T}$ at level $\ell$. In particular, $\ssb_{-1}$ and $\ssb_{n-1}$ identify the initial node 
and the final node on trellis $\const{T}$ at levels $-1$ and $n-1$, respectively.
In addition, the branch connecting state $\ssb_{\ell-1}$ and state $\ssb_\ell$
is labeled with code bit $v_\ell$.
As such, the one-to-one mapping between
codewords of \C\ and paths over $\const{T}$
is built. This completes the construction of trellis $\const{T}$ based on parity-check matrix $\Hm$.
The super-trellis $\overline{\const{T}}$
corresponding to supercode $\overline\C$ and its parity-check matrix $\overline\Hm$ can be similarly constructed, of which its state at level $\ell$ is denoted by $\bar\ssb_\ell$.

We next introduce the ML soft-decision decoding for codes
with trellis representation.
Denote again by $\vb\triangleq(v_0,v_1,\ldots,v_{n-1})$ a binary zero-one
codeword of $\C$.
Define the hard-decision sequence $\yb=(y_0,y_1,\ldots,y_{n-1})$ corresponding to the received vector $\rb=(r_0,r_1,\ldots,r_{n-1})$ as
$$
y_j\triangleq\begin{cases}
1,&\text{if}\ \phi_j<0;\\
0,&\text{otherwise,}
\end{cases}
$$
where
$$
\phi_j\triangleq\log {{\Pr(r_j|0})\over \Pr(r_j|1)}
$$
is the log-likelihood ratio, 
and $\Pr(r_j|0)$ and $\Pr(r_j|1)$ are the conditional probabilities of receiving $r_j$ given $0$ and $1$ were transmitted, respectively. Here, $\Pr(r_j|0)$ can be either a probability density function (pdf) for continuous (unquantized) $r_j$ or a probability mass function (pmf) for discrete (softly quantized) $r_j$.

The syndrome of $\yb$ is given by $\yb\Hm^\Top$, where  
superscript ``$\Top$" denotes the matrix transpose operation. Let $E(\ab)$ be the
collection of all error patterns whose syndrome is $\ab$. Then,
the maximum-likelihood (ML) decoding output $\hat{\vb}$ for
received vector $\rb$ satisfies:
$$\hat{\vb}=\yb\oplus\eb^\ast,$$
where $\eb^\ast=(e_0^\ast,e_1^\ast,\ldots,e_{n-1}^\ast)\in E(\yb\Hm^\Top)$ is the error pattern
satisfying
$$
\sum^{n-1}_{j=0}\
e^{\ast}_j|\phi_j|\le \sum^{n-1}_{j=0} \ e_{j}|\phi_j|$$
for all $\eb=(e_0,e_1,\ldots,e_{n-1})\in  E(\yb\Hm^\Top)$, 
and ``$\oplus$" denotes component-wise modulo-two addition.
We thereby define
a new metric for paths in a trellis as follows.

\medskip

\begin{definition}[ML path metric]\label{defi1}
For a path with labels
$\xb_{(\ell)}=(x_0, x_1,\ldots, x_{\ell})$,
which ends at level $\ell$ on trellis $\const{T}$,
define the {\it metric} associated with it as
$$
M\left(\xb_{(\ell)}\right)\triangleq\sum^{\ell}_{j=0} M(x_j),
$$
where $M(x_j)\triangleq(y_j\oplus x_j)|\phi_j|$ is the
{\it bit metric}. 
Similarly, for a backward path with labels
$\bar\xb_{[\ell]}=(\bar x_{\ell},\bar x_{\ell+1},\ldots,\bar x_{n-1})$
on super-trellis $\overline{\const{T}}$,
define the {\it metric} associated with it as
\begin{equation}
\label{b-metric}
M\left(\bar\xb_{[\ell]}\right)\triangleq\sum^{n-1}_{j=\ell} M(\bar x_j).
\end{equation}
\end{definition}

After giving the notions of supercode  and super-trellis as well as path metrics, we proceed to present the proposed two-phase decoding scheme in the next section.

\section{Two-phase ML Soft-Decision Decoding Algorithm for Binary Linear Block Codes}
\label{sec:proposed}

As mentioned in the introduction section, the proposed decoding algorithm  has two phases.

The first phase applies the Viterbi algorithm backwardly
to the supe-trellis derived from the parity-check matrix $\overline\Hm$ of supercode $\overline\C$ using 
the path metric defined in \eqref{b-metric}, during which the path metric of the backward survivor starting from the final node at level $n-1$
and ending at a node corresponding to state $\bar\ssb_\ell$ at level $\ell$ is retained for use in the second phase.
For convenience of referring it, we denote this path metric by $c(\bar\ssb_\ell)$.
At the end of the first phase, a backward survivor path ending at the initial node at level $-1$ is resulted. 
The backward Viterbi algorithm in the first phase is summarized below.

\medskip

\noindent
$\langle$Phase 1: The backward Viterbi Algorithm$\rangle$
\begin{list}{Step~\arabic{step}.}
    {\usecounter{step}
    \setlength{\labelwidth}{1cm}
    \setlength{\leftmargin}{1.6cm}\slshape}
\item Associate zero initial metric
    with the backward path\footnote{It is clear that a path on a trellis can not only be identified by its labels, but also be determined by the states it traverses. Accordingly, 
 path $\bar\xb_{[\ell]}$ can be equivalently designated by $(\bar\ssb_\ell,\bar\ssb_{\ell+1},\ldots,\bar\ssb_{n-1})$.
    } containing only the final state $\bar\ssb_{n-1}$ on super-trellis $\overline{\const{T}}$, and let $c(\bar\ssb_{n-1})=0$. Set $\ell=n-1$.
\item Decrease $\ell$ by one. Compute the metrics for all backward paths
extending from the backward survivors ending at level $\ell+1$
(and hence entering a state at level $\ell$). For each state $\bar\ssb_\ell$ at level $\ell$,
    keep the entering path with the least metric as its survisor, and delete the remaining.
    Let $c(\bar\ssb_\ell)$ be this least metric.
   \label{VA_loop}
\item If $\ell=0$, stop the algorithm; otherwise, go to Step~\ref{VA_loop}.
\end{list}

\medskip

In the second phase, the priority-first search algorithm
is operated on trellis $\const{T}$ in the usual forward fashion (i.e., from level $0$ to level $n-1$);
hence, the second phase always outputs a codeword in $\C$.

Now for each path with labels $\xb_{(\ell)}$ on trellis $\const{T}$, an evaluation
function $f$ associated with it is defined as:
$$
f\left(\xb_{(\ell)}\right)=g\left(\xb_{(\ell)}\right)+h\left(\xb_{(\ell)}\right),
$$
where 
the value of $g$-function is assigned according to:
\begin{equation}
\label{fun-g}
g\left(\xb_{(\ell)}\right)=\begin{cases}
0,&\ell=-1;\\
g\left(\xb_{(\ell-1)}\right)+M(x_{(\ell)}),&\ell=0,1,\ldots,n-1
\end{cases}
\end{equation}
and the value of $h$-function is given by:
\begin{equation}
\label{fun-h}
h\left(\xb_{(\ell)}\right)=c\left(\beta(\ssb_{\ell})\right).
\end{equation}
In \eqref{fun-h}, $\ssb_\ell$ is the ending state of the path with label $\xb_{(\ell)}$, 
and $\beta(\ssb_{\ell})$ is the state $\bar\ssb_\ell$ on super-trellis $\overline{\const{T}}$ that has the same first $(n-\bar k)$ components
as $\ssb_\ell$. Note that $\beta(\ssb_\ell)$ exists and is well-defined 
for every $\ssb_\ell$ on trellis $\const{T}$ since the parity-check matrices 
of $\C$ and $\overline\C$ satisfy \eqref{eq:H}.

It can be verified that
$f(\xb_{(n-1)})=g(\xb_{(n-1)})$
since $\beta(\ssb_{n-1})={\bf 0}$ and $h\left(\xb_{(n-1)}\right)=c(\beta(\ssb_{n-1}))=0$. 
This implies that the path with the minimum $f$-function value on trellis $\const{T}$ is
exactly the one with the minimum ML path metric.

Two storage spaces are necessary for the priority-first search over trellis $\const{T}$.
The  {\em Open Stack} records the paths visited thus far by the priority-first search,
while the {\em Close Table} keeps the starting and ending states and ending levels of the paths that have ever been on top of the Open Stack. They are so named because
the paths in  the Open Stack can be further extended and hence remain {\em open}, but
the paths with information in the Closed Table are {\em closed} for further extension.

We summarize the priority-first search  algorithm  over trellis $\const{T}$ in the following.

\medskip

\noindent
$\langle$Phase 2: The Priority-First Search Algorithm$\rangle$
\begin{list}{Step~\arabic{step}.}
    {\usecounter{step}
    \setlength{\labelwidth}{1cm}
    \setlength{\leftmargin}{1.6cm}\slshape}
\item Let $\rho=\infty$,  and assign $\xb=\emptyset$. 
\label{PF_rho}

\item Load into the Open Stack the path
containing only the initial state $\ssb_{-1}$ at level $-1$. \label{PF_ini}

\item If the Open Stack is empty, output $\xb$
      as the final ML decision, and stop the algorithm.\label{PF_loop}
      
\item If the starting and ending states and ending level of the top path in the Open Stack have
      been recorded in the Close Table, discard the top path from the Open Stack,
      and go to Step~\ref{PF_loop}; otherwise, record the starting
      and ending states and ending level of this top path in the Close Table.\label{PF_dis}
\item Compute the
      $f$-function values of the successors of the top path in the Open Stack,
      and delete the top path from the Open Stack.
      If the $f$-function value of any successor
      is equal to or greater than $\rho$, just delete it.
      \label{PF_del}
\item For all remaining successor paths that reach level $n-1$,
set $\rho$ to be the least path metric among them, and
update $\xb$ as the successor path corresponding to this least path metric and discard all the others.
\label{PF_ub}
\item Insert the remaining successor paths (from Steps \ref{PF_del} and \ref{PF_ub}) into the Open Stack,
      and re-order the paths in the Open Stack according to
      ascending $f$-function values.
      Go to Step~\ref{PF_loop}.
      \label{PF_last}
\end{list}


\section{Optimality of the Proposed Algorithm}
\label{sec:optimality}

This section proves the optimality of 
the proposed two-phase decoding algorithm.
We begin with two essential lemmas required for the optimality proof.

\medskip

\begin{lemma}
\label{branch}
Let path $\xb_{(\ell+1)}$ 
be an immediate successor of
path $\xb_{(\ell)}$ on trellis $\const{T}$.
Denote the ending states of $\xb_{(\ell+1)}$ and $\xb_{(\ell)}$ by 
$\ssb_{\ell+1}$ and $\ssb_\ell$, respectively.
Then,
$$\beta(\ssb_{\ell+1})=\beta(\ssb_{\ell})+x_{\ell+1}\bar\hb_{\ell+1},$$
where $\bar\hb_{\ell+1}$ is the $(\ell+2)$th column of parity-check matrix $\overline\Hm$.
In other words, there exists a branch between $\beta(\ssb_{\ell})$ and $\beta(\ssb_{\ell+1})$ with label $x_{\ell+1}$ over super-trellis $\overline{\const{T}}$.
\end{lemma}
\begin{IEEEproof}
Recall that $$\ssb_{\ell+1}=\ssb_{\ell}+x_{\ell+1}\hb_{\ell+1}$$ and
$$\hb_{\ell+1}=\begin{bmatrix}
\ \bar\hb_{\ell+1}\ \\ \ \pb_{\ell+1}\
\end{bmatrix}$$
for some $\pb_{\ell+1}$ according to \eqref{eq:H}. 
It is thus obvious that
$$\beta(\ssb_{\ell+1})=\beta(\ssb_{\ell})+x_{\ell+1}\bar\hb_{\ell+1}$$
since $\bar\hb_{\ell+1}$ contains the first $(n-\bar k)$ components of $\hb_{\ell+1}$.
\end{IEEEproof}

\medskip

\begin{lemma}
\label{non-decreasing}
$f$ is a non-decreasing function along any  path on trellis $\const{T}$, i.e.,
$$f\left(\xb_{(\ell)}\right)
\le f\left(\xb_{(\ell+1)}\right),$$
where path $\xb_{(\ell+1)}$
is an immediate successor of
path $\xb_{(\ell)}$ over trellis $\const{T}$.
\end{lemma}
\begin{IEEEproof} 
The fundamental attribute of the backward Viterbi algorithm in the first phase
gives that $c(\beta(\ssb_\ell))$ is the minimum metric
among all backward paths that end at state $\beta(\ssb_\ell)$ at level $\ell$.
By Lemma~\ref{branch},
we have:
$$c(\beta(\ssb_\ell))\leq c(\beta(\ssb_{\ell+1}))+ M(x_{\ell+1}),$$
where $\ssb_{\ell+1}$ and $\ssb_\ell$ are respectively the states
that paths $\xb_{(\ell+1)}$ and $\xb_{(\ell)}$ end at.
Hence, we derive:
\begin{eqnarray*}
 f\left(\xb_{(\ell+1)}\right)
 &=&g\left(\xb_{(\ell+1)}\right)
 +h\left(\xb_{(\ell+1)}\right)\\
 &=&g\left(\xb_{(\ell)}\right)
 + M(x_{\ell+1})
 +c(\beta(\ssb_{\ell+1}))\\
 &\ge&g\left(\xb_{(\ell)}\right)+c(\beta(\ssb_{\ell}))\\
 &=&f\left(\xb_{(\ell)}\right).
 \end{eqnarray*}
\end{IEEEproof}

\medskip

Based on these two lemmas, the next theorem proves the optimality of the proposed two-phase algorithm.

\medskip

\begin{theorem}
In the second phase, the priority-first search algorithm always output an ML path.
\end{theorem}
\begin{IEEEproof}
It suffices to prove that if the Open Stack is empty, the algorithm will output
an ML path as claimed in Step~\ref{PF_loop}.
This can be confirmed by showing that 
Steps~\ref{PF_dis} and~\ref{PF_del}
never delete any ML path.

Suppose that in Step~\ref{PF_dis}, the starting
and ending states and ending level of the new top
path $\xb_{(\ell)}$ have been recorded
in the Close Table at some previous time due to path
$\hat{\xb}_{(\ell)}$.
Since path $\xb_{(\ell)}$ must be an offspring of some path
$\xb_{(j)}$ that once
coexisted with path $\hat{\xb}_{(\ell)}$ in the Open Stack
at the time path $\hat{\xb}_{(\ell)}$ was on top of
the Open Stack,  where $j<\ell$, we have
\begin{equation}
\label{ineq}
f\left(\xb_{(\ell)}\right)\geq
f\left(\xb_{(j)}\right)\geq f\left(\hat{\xb}_{(\ell)}\right).
\end{equation}
Notably, the first inequality in \eqref{ineq} follows from
Lemma~\ref{non-decreasing}, and the second inequality in \eqref{ineq} is valid
because the top path in the Open Stack always carries the minimum $f$-function value among all coexisting paths.
As a result, the offsprings of path $\xb_{(\ell)}$ ending at level $n-1$
cannot yield smaller metrics than those length-$n$ offsprings of path $\hat{\xb}_{(\ell)}$, and hence deletion of path $\xb_{(\ell)}$ will not compromise the optimality of the decoding algorithm.

For Step~\ref{PF_del}, we argue that $\rho$ is either a trivial upper bound of the
final ML path (cf.~Step~\ref{PF_rho}) or the metric of a valid path
that reaches level $n-1$ (cf.~Step~\ref{PF_ub}), so deletion of any successor paths
whose $f$-function values are no less than $\rho$
will never eliminate any ML path. 
This completes the proof of optimality of the proposed algorithm.
\end{IEEEproof}

\section{Evaluation of Computational Efforts}
\label{sec:performance}

In this section, we investigate by simulations
the computational effort of the
proposed decoding algorithm over the additive white Gaussian noise (AWGN) channels.
We assume that the codeword is antipodally modulated, and hence
the received vector is given by
$$r_j=(-1)^{v_j}\sqrt{\cal E}+\lambda_j,$$
for $0\leq j\leq n-1$, where ${\cal E}$ is the signal energy per
channel bit, and $\{\lambda_j\}_{j=0}^{n-1}$ are independent noise
samples of a white Gaussian process with single-sided noise power
per hertz $N_0$. The signal-to-noise ratio (SNR) for the channel
is therefore given by $\text{SNR}\triangleq {\cal E}/N_0$. In order to account
for the code redundancy for different code rates, we will use the
SNR per information bit in the following discussion, which is defined as
$$\text{SNR}_{\text{b}}=\frac{n{\cal E}/k}{N_0}=\frac{n}{k}
\left(\frac{\cal E}{N_0}\right).$$

It can be easily verified that for antipodal-input AWGN channels, the log-likelihood ratio $\phi_j$ is a fixed multiple 
of the received scalar $r_j$; thus, the metric
associated with a path $\xb_{(\ell)}$ can be equivalently simplified to
$$
M\left(\xb_{(\ell)}\right)\triangleq\sum^{\ell}_{j=0}
(y_j\oplus x_j)|r_j|,
$$
where
$$
y_j\triangleq
\begin{cases}
1,&\text{if}\ r_j<0;\\
0,&\text{otherwise.}
\end{cases}
$$

The decoding complexity in the first phase is clearly determined by the number of bit metric computations performed. We emphasize that the decoding complexity
in the second phase can also be regulated by the number of $f$-function evaluations (equivalently, the number of bit metric computations as indicated in \eqref{fun-g}) during 
the priority-first search.
This is due to that the cost of searching and re-ordering of stack elements
can be made a constant multiple of the computational complexity by adopting 
a priority-queue data structure  
in stack implementation \cite{COR91}. One can even employ a hardware-based stack structure \cite{LAV94} and attain constant complexity in stack maintenance.
Therefore, to use the number of overall metric computations
as the key determinant of algorithmic complexity for our proposed two-phase decoding algorithm is justified.

We now turn to empirical examination of the average decoding
complexity of the proposed tpMLSD algorithm.
The linear block code considered is the $r$th order binary  Reed-Muller code, RM$(r,m)$,
which is an $(n,k)$ linear block code with $n=2^m$ and $k=1+\sum_{i=1}^r{m\choose i}$.
It is known~\cite{MAC77} that RM$(r+i,m)$ is a supercode of RM$(r,m)$ for $i\geq 1$. 
In our simulations, $\C$ is RM$(2,6)$ and $\overline\C$ is RM$(4,6)$;
hence, $n=64$, $k=22$ and $\bar k=57$. 
Under the same optimal ML performance, we compare the proposed two-phase ML soft-decision decoding (tpMLSD) algorithm with the recursive ML decoding (RMLD) algorithm~\cite{FUJ98} and the list ML decoding (LMLD) algorithm~\cite{FRE04} in average decoding complexity, and summarize the results in Table~\ref{tab:complexity}. 

Note that instead of listing the decoding complexity of the LMLD,
lower bounds obtained from decoding its supercode counterpart 
using the marking algorithm are given~\cite{FRE04}. 
Apparently, the real decoding complexity of the LMLD is higher than this lower bound. 
The table then shows that the LMLD is much more complex 
than the other two algorithms,
and our two-phase decoding algorithm
consumes only 1/13 of the computational effort of the RMLD at $\text{SNR}_{\text{b}} = 4.5$ dB, in which circumstance the bit error rate (BER) is around $10^{-5}$.
Further, when $\text{SNR}_{\text{b}}$ is reduced to $3$ dB, the average computational complexity of the proposed two-phase decoding scheme can still reach $1/8$ of that of the RMLD. 

\begin{table}
\caption{Average computational complexities (I.e., average number of metrics evaluated) of the RMLD, the LMLD, and the tpMLSD.
The linear block code considered is RM$(2,6)$, while
the suppercode used in the tpMLSD is RM$(4,6)$.}
\label{tab:complexity}
\begin{center}
\begin{tabular}{|c||c|c|c|c|c|}\hline
$\text{SNR}_{\text{b}}$ & 3 dB & 3.5 dB & 4 dB & 4.5 dB & 5 dB\\ \hline\hline
RMLD~\cite{FUJ98} & 78209 & 78209 & 78209 & 78209 & 78209 \\ \hline
*LMLD~\cite{FRE04} &  2097152 &  2097152 & 2097152 &  2097152&  2097152 \\ \hline
tpMLSD & 10078 & 7863 & 6602 & 6010 & 5695 \\ \hline
\end{tabular}
\end{center}
*What are listed here are lower bounds to the decoding complexities of the LMLD.
\end{table}

\section{Conclusion}
\label{sec:conclusion}

In this work, we proposed a two-phase scheme for ML soft-decision decoding of linear block codes. This novel decoding algorithm has two phases, where the backward Viterbi algorithm is employed on a supercode of the linear block code in the first phase, while the priority-first search algorithm is performed on the trellis of the linear block code in the second phase. Simulations showed that the computational complexity of the proposed two-phase scheme 
is one order of magnitude better than that of the RMLD when $\text{SNR}_{\text{b}} = 4.5$ dB. 
Since such a new approach can be extended to decoding any linear block codes
when their supercodes are obtainable,
a possible future work 
is to extend this two-phase decoding scheme to codes like Reed-Solomon,
for which maximum-likelihood soft-decision decoding 
is generally considered a challenging task.

\bibliographystyle{ieeebib}
\bibliography{codingn}

\end{document}